\documentclass[12pt]{article}
\usepackage{epsfig}
\usepackage{amsmath}
\usepackage{hhline}
\usepackage{amssymb}
\usepackage{times}
\usepackage{color}
\usepackage{cite}
\usepackage{multirow}
\usepackage[]{lineno}

\newlength{\dinwidth}
\newlength{\dinmargin}
\def\PLB{{\em Phys. Lett.}   {\bf B}}

\begin{document}

\begin{titlepage}
\noindent

\begin{flushleft}
{\tt DESY 11-123\hfill ISSN 0418-9833} \\
{\tt July 2011}
\end{flushleft}

\vspace*{3cm}

\begin{center}
\begin{Large}

{\bf Search for First Generation Leptoquarks in {\boldmath{$ep$}} Collisions at HERA} 

\vspace{2cm}

H1 Collaboration

\end{Large}
\end{center}

\vspace{2cm}

\begin{abstract} \noindent

A search for first generation scalar and vector leptoquarks produced
in $ep$ collisions is performed by the H1 experiment at HERA.
The full H1 data sample is used in the analysis, corresponding to an
integrated luminosity of $446$~pb$^{-1}$.
No evidence for the production of leptoquarks is observed in
final states with a large transverse momentum electron or with
large missing transverse momentum, and constraints on leptoquark
models are derived.
For leptoquark couplings of electromagnetic strength $\lambda=0.3$,
first generation leptoquarks with masses up to $800$~GeV are excluded
at $95\%$ confidence level.

\noindent
\end{abstract}

\vspace{5mm}
\begin{center}
Accepted by \PLB
\end{center}

\end{titlepage}

\begin{flushleft}
F.D.~Aaron$^{5,48}$,           
C.~Alexa$^{5}$,                
V.~Andreev$^{25}$,             
S.~Backovic$^{30}$,            
A.~Baghdasaryan$^{38}$,        
S.~Baghdasaryan$^{38}$,        
E.~Barrelet$^{29}$,            
W.~Bartel$^{11}$,              
K.~Begzsuren$^{35}$,           
A.~Belousov$^{25}$
% ...

\bigskip{\it
 $ ^{1}$ I. Physikalisches Institut der RWTH, Aachen, Germany \\
% ...
$ ^{5}$ National Institute for Physics and Nuclear Engineering (NIPNE) ,
          Bucharest, Romania$^{ m}$ \\
% ...
$ ^{11}$ DESY, Hamburg, Germany \\
% ...
$ ^{25}$ Lebedev Physical Institute, Moscow, Russia$^{ e}$ \\
% ...
$ ^{29}$ LPNHE, Universit\'e Pierre et Marie Curie Paris 6,
          Universit\'e Denis Diderot Paris 7, CNRS/IN2P3, Paris, France \\
 $ ^{30}$ Faculty of Science, University of Montenegro,
          Podgorica, Montenegro$^{ n}$ \\
% ...
$ ^{35}$ Institute of Physics and Technology of the Mongolian
          Academy of Sciences, Ulaanbaatar, Mongolia \\
% ...
$ ^{38}$ Yerevan Physics Institute, Yerevan, Armenia \\
% ...
$ ^{48}$ Also at Faculty of Physics, University of Bucharest,
          Bucharest, Romania \\
% ...
$ ^e$ Supported by the Deutsche Forschungsgemeinschaft \\
% ...
$ ^m$ Supported by the Romanian National Authority for Scientific Research
      under the contract PN 09370101 \\
 $ ^n$ Partially Supported by Ministry of Science of Montenegro,
      no. 05-1/3-3352 \\
}
\end{flushleft}

\newpage
\section{Introduction}
\label{sec:intro}

The $ep$ collisions at HERA provide a unique opportunity to search for new
particles coupling directly to a lepton and a quark.
An example of such particles are leptoquarks (LQs), colour triplet bosons
which appear in a variety of beyond the Standard Model (SM)
theories~\cite{Pati:1974yy,Schrempp:1984nj,Dimopoulos:1979es,Nilles:1983ge}.
Particle interactions in the SM conserve lepton flavour.
If this property is extended to LQ models any such particles produced at HERA would
decay exclusively into a quark and a first generation lepton, namely an
electron\footnote{In this paper the term ``electron'' is used generically to refer to both
electrons and positrons, if not otherwise stated.} or a neutrino.
Searches for such signatures have previously been performed
at HERA~\cite{Aktas:2005pr,Adloff:1999tp,Chekanov:2003af}.
A dedicated analysis investigating the production of second and third generation
leptoquarks has also been performed by the H1 Collaboration, where the final state
contains a muon or the decay products of a tau lepton in combination with a
hadronic system~\cite{h1lfv2011}.


In this paper a search for leptoquarks coupling exclusively to a quark and a first
generation lepton is performed using the full $e^{\pm}p$ collision data set taken
in the years 1994-2007 by the H1 experiment at HERA.
The data were recorded with an electron beam of energy $27.6$~GeV, which was
longitudinally polarised up to $38$\%, and a proton beam of energy up to
$920$~GeV, corresponding to a centre-of-mass energy $\sqrt{s}$ of up to $319$~GeV.
The total integrated luminosity of the analysed data is $446$~pb$^{-1}$, which
represents a factor of almost four increase with respect to the previously
published H1 results.
The presented results supersede those derived in previous searches for first generation
leptoquarks by the H1 Collaboration~\cite{Aktas:2005pr}.

\section{Leptoquark Phenomenology and Standard Model Processes}
\label{sec:theory}

\subsection{Leptoquark production at HERA}
\label{sec:lqs}

The phenomenology of LQs at HERA is discussed in detail elsewhere~\cite{Adloff:1999tp}.
The effective Lagrangian considered conserves lepton and baryon number, obeys
the symmetries of the SM gauge groups
${\rm SU}(2)_L \times {\rm U}(1)_Y$ and ${\rm SU}(3)_C$
and includes both scalar and vector LQs.
In the framework of the phenomenological Buchm\"uller-R\"uckl-Wyler (BRW)
model~\cite{BRW}, LQs are classified into $14$ types~\cite{14LQ} with respect to the
quantum numbers spin $J$, weak isospin $I$ and chirality $C$ (left-handed $L$,
right-handed $R$).
Scalar ($J=0$) LQs are denoted as $S_{I}^{C}$ and vector ($J=1$) LQs are denoted
$V_{I}^{C}$ in the following.
LQs with identical quantum numbers except for weak hypercharge are distinguished
using a tilde, for example $V_0^R$ and $\tilde{V}_0^R$.
Whereas all $14$ LQs couple to electron-quark pairs, four of the left-handed LQs,
namely $S_0^L$, $S_{1}^L$, $V_0^L$ and $V_{1}^L$, may also decay to a
neutrino-quark pair.
In particular, for $S_0^L$ and $V_0^L$ the branching fraction of decays into an
electron-quark pair is predicted by the model to be
\mbox{$\beta_e\!=\!\Gamma_{\rm eq}/(\Gamma_{\rm eq}+\Gamma_{\rm \nu q})\!=0.5$},
where \mbox{$\Gamma_{\rm eq}$} (\mbox{$\Gamma_{\rm \nu q}$}) denotes the partial
width for the LQ decay to an electron (neutrino) and a quark $q$.
The branching fraction of decays into a neutrino-quark pair is then given by
$\beta_{\nu} =1-\beta_{e}$.


Leptoquarks carry both lepton ($L$) and baryon ($B$) quantum numbers.
The fermion number \mbox{$F\!=\!L\!+\!3\,B$} is conserved.
Leptoquark processes in $ep$ collisions proceed directly via $s$-channel
resonant LQ production or indirectly via $u$-channel virtual LQ exchange.
A dimensionless parameter $\lambda$ defines the coupling at the
lepton-quark-LQ vertex.
For LQ masses $M_{\rm LQ}$ below $\sqrt{s}$, the $s$-channel production of
$F = 2$ ($F = 0$) LQs dominates in $e^-p$ ($e^+p$) collisions.
For LQ masses above $\sqrt{s}$, both the $s$ and $u$-channel, as well
as the interference with SM processes, are important such that both $e^-p$ and
$e^+p$ collisions have similar sensitivity to all LQ types.

\subsection{Standard Model processes}
\label{sec:sm}

The search reported here considers final states where the leptoquark decays into an
electron and a quark or a neutrino and a quark.
Such decays lead to topologies similar to those of deep-inelastic scattering
(DIS) neutral current (NC) and charged current (CC) interactions at high negative
four-momentum transfer squared $Q^2$.
The analysis is therefore performed using event selections (see
section~\ref{sec:sel}) similar to those used in inclusive DIS
analyses~\cite{Adloff:2003uh} and previous LQ searches~\cite{Aktas:2005pr}.


The SM prediction for both NC and CC DIS processes is obtained using the Monte Carlo (MC)
event generator DJANGOH~\cite{django}, which is based on LEPTO~\cite{lepto} for the hard
interaction and HERACLES~\cite{heracles} for leptonic single photon emission and virtual
electroweak corrections.
LEPTO combines $\mathcal{O}(\alpha_s)$ matrix elements with higher order QCD effects
using the colour dipole model as implemented in ARIADNE~\cite{ariadne}.
The JETSET program~\cite{jetset} is used to simulate the hadronisation process.
Additional SM background contributions from photoproduction processes are simulated
using the PYTHIA~\cite{Sjostrand:2000wi} event generator, with the GRV-G~LO~\cite{grv}
parameterisation of the photon parton density functions (PDFs).
All SM expectations are based on the CTEQ6m~\cite{Pumplin:2002vw} proton parton
density function parameterisation, which includes only $12$\% of
the H1 data analysed in this paper, in addition to $30$~pb$^{-1}$ of
ZEUS data. At high Bjorken $x$, the CTEQ6m parameterisations
are dominated by data from fixed target experiments due to the
limited amount of HERA data included.


Generated events are passed through a GEANT~\cite{Brun:1987ma} based simulation
of the H1 apparatus,  which takes into account the running conditions of the data taking.
Simulated events are reconstructed and analysed using the same program chain as is used
for the data.

\section{Experimental Method}
\label{sec:exp}

\subsection{Data sets and lepton polarisation}
\label{sec:datapol}

The full H1 data sample is made up of $164$~pb$^{-1}$ recorded in $e^{-}p$ collisions
and $282$~pb$^{-1}$ in $e^{+}p$ collisions, of which $35$~pb$^{-1}$
were recorded at $\sqrt{s} = 301$~GeV.
Data collected from 2003 onwards were taken with a longitudinally polarised lepton beam.
As leptoquarks are chiral particles, these data are analysed in separate polarisation samples,
formed by combining all data periods with similar lepton beam polarisation
$P_{e} = (N_{R} - N_{L})/(N_{R} + N_{L})$, where $N_{R}$ ($N_{L}$) is the number
of right- (left-)
handed leptons in the beam.
The average polarisation and luminosity of all data sets are detailed in
table~\ref{tab:alldatasets}.

\begin{table}[h]
  \renewcommand{\arraystretch}{1.2}
 \begin{center}
  \begin{tabular}{|c||c|c|c|}
  \hline
 Collisions & $\sqrt{s}$~~[GeV] & $P_{e}$~~[\%] & $\mathcal{L}$~~[pb$^{-1}$]\\
   \hline
   \hline
   $e^{+}p$ &  $301$ & $0$ & $35$ \\
   $e^{-}p$ &  $319$ & $0$ & $15$ \\
   $e^{+}p$ &  $319$ & $0$ & $67$ \\
   \hline
   $e^{+}p$ &  $319$ & $+32$ & $98$ \\
   $e^{+}p$ &  $319$ & $-38$ & $82$ \\
   $e^{-}p$ &  $319$ & $+37$ & $46$ \\
   $e^{-}p$ &  $319$ & $-26$ & $103$ \\
   \hline
\end{tabular}
 \end{center}
  \caption{Centre-of-mass energy $\sqrt{s}$, average lepton beam polarisation $P_{e}$
    and integrated luminosity $\mathcal{L}$ of the analysed H1 data sets.}
 \label{tab:alldatasets}
\end{table}

\subsection{H1 detector}
\label{sec:det}

A detailed description of the H1 experiment can be found elsewhere~\cite{Abt:h1}.
Only the detector components relevant to this analysis are briefly described here.
A right-handed Cartesian coordinate system is used with the origin at the nominal primary
$ep$ interaction vertex. 
The proton beam direction defines the positive $z$ axis (forward direction).
The polar angle $\theta$ and the transverse momenta $P_T$ of all particles are defined with
respect to this axis.
The azimuthal angle $\phi$ defines the particle direction in the transverse plane.


The Liquid Argon (LAr) calorimeter~\cite{Andrieu:1993kh} covers the polar angle range
\mbox{$4^\circ < \theta < 154^\circ$} with full azimuthal acceptance.
The energies of electromagnetic showers are measured in the LAr calorimeter with a precision of
\mbox{$\sigma (E)/E \simeq 11\%/ \sqrt{E/\mbox{GeV}} \oplus 1\%$} and hadronic
energy deposits with \mbox{$\sigma (E)/E \simeq 50\%/\sqrt{E/\mbox{GeV}} \oplus 2\%$},
as determined in test beam measurements~\cite{Andrieu:1993tz,Andrieu:1994yn}.
A lead-scintillating fibre calorimeter\footnote{This device was installed in 1995, replacing a
lead-scintillator sandwich calorimeter~\cite{Abt:h1}.} (SpaCal)~\cite{SpaCal} covering the
backward region $153^\circ < \theta < 178^\circ$ completes the measurement of
charged and neutral particles.
The central \mbox{($20^\circ < \theta < 160^\circ$)}  and forward
\mbox{($7^\circ < \theta < 25^\circ$)}  inner tracking detectors are used to measure charged
particle trajectories and to reconstruct the interaction vertex.
The measured trajectories fitted to the interaction vertex are referred to as tracks
in the following.
The LAr calorimeter and inner tracking detectors are enclosed in a superconducting
magnetic coil with a field strength of $1.16$~T.
From the curvature of charged particle trajectories in the magnetic field, the central
tracking system provides transverse momentum measurements with a resolution of 
\mbox{$\sigma_{P_T}/P_T = 0.002 P_T / \mbox{GeV}$} $\oplus~0.015$.
The return yoke of the magnetic coil is the outermost part of the detector and is equipped with
streamer tubes forming the central muon detector \mbox{($4^\circ < \theta < 171^\circ$)}. 
In the very forward region of the detector \mbox{($3^\circ < \theta < 17^\circ$)} a set of drift
chambers detects muons and measures their momenta using an iron toroidal magnet.
The luminosity is determined from the rate of the Bethe-Heitler process
$ep \rightarrow ep \gamma$, measured using a photon detector located close to the beam
pipe at $z=-103~{\rm m}$, in the backward direction.


The LAr calorimeter provides the main trigger in this analysis~\cite{Adloff:2003uh}.
In order to remove events induced by cosmic rays and other non-$ep$ background,
the event vertex is required to be reconstructed within $\pm 35$~cm in $z$ of the average
nominal interaction point.
In addition, topological filters and timing vetoes are applied.

\subsection{Particle identification and event reconstruction}
\label{sec:partid}

The scattered electron is identified as a compact and isolated cluster of energy in the
electromagnetic part of the LAr calorimeter with an associated track in the inner tracking
detectors.
The hadronic final state is reconstructed using a particle flow algorithm to combine
tracks and calorimeter deposits not associated to the scattered electron~\cite{h1lt,roman}.
The missing transverse momentum $P_{T}^{\rm miss}$, which may indicate the presence
of neutrinos in the final state, is derived from all reconstructed particles in the event.


The kinematic quantities in NC events are determined using the electron
method~\cite{eandhadronmethod}, which uses information exclusively from the
scattered electron.
In CC events, the kinematic quantities are determined exclusively from the hadronic
final state~\cite{eandhadronmethod}.
The leptoquark mass $M_{\rm LQ}~=~\sqrt{Q^2/y}$ is reconstructed using the measured
kinematics of the scattered electron~(had-ronic final state) in the analysis of NC (CC)
topologies, where $y$ is the inelasticity.

\section{Data Analysis}
\label{sec:analysis}

\subsection{DIS event selections}
\label{sec:sel}

Neutral current events are selected by requiring a scattered electron with energy
$E_{e} > 11$~GeV and $Q^{2} > 133$~GeV$^2$.
Additionally, a kinematic cut on the inelasticity $0.1~<~y<~0.9$ is employed to
remove regions of poor reconstruction, poor resolution, large QED radiative effects
and background from photoproduction processes~\cite{Aktas:2005pr}.
Background from neutral hadrons or photons misidentified as leptons is suppressed
by requiring a charged track to be associated to the lepton candidate.
Energy-momentum conservation requires that
$\Sigma_{i} (E^{i} - P^{i}_{z}) = 2E^{0}_{e}$, where the sum runs over all reconstructed
particles, $P_{z}$ is the momentum along the proton beam axis and $E^{0}_{e}$ is the
electron beam energy.
Applying the condition $\Sigma_{i} (E^{i} - P^{i}_{z})  > 35$~GeV further suppresses the
contamination from photoproduction background in which the scattered lepton is
undetected in the backward direction and a hadron is misidentified as an electron.
The $\Sigma_{i} (E^{i} - P^{i}_{z})$ requirement also further suppresses the influence
of radiative corrections arising from initial state bremsstrahlung.


Charged current events are selected by requiring significant missing transverse
momentum, $P_{T}^{\rm miss}>12$~GeV, which is due to the undetected neutrino.
To ensure a high trigger efficiency and good kinematic resolution, the analysis is
further restricted to the region $0.1 <  y < 0.85$.
The main SM background is due to photoproduction events, in which the  scattered
electron escapes undetected in the backward direction and transverse momentum is
missing due to fluctuations in the detector response or undetected particles.
This background is suppressed by exploiting the correlation between $P_{T}^{\rm miss}$
and the ratio $V_{\rm ap} / V_{\rm p}$~\cite{Adloff:1999tp} of transverse energy flow
anti-parallel and parallel to the hadronic final state transverse momentum
vector~\cite{ringaile}.

\subsection{Systematic uncertainties}
\label{sec:sys}

The experimental systematic uncertainties included in the analysis of the polarised
data taken in the years 2003-2007 are described in the following.
The systematic uncertainties on the 1994-2000 data are described in the previous
H1 publication~\cite{Aktas:2005pr}.


In the NC event samples, a systematic scale uncertainty of $1$-$3$\% is assigned
to the electromagnetic energy measured in the LAr calorimeter, depending on the
$z$-coordinate of the impact position of the scattered electron.
A $0.5$\% component of this uncertainty is considered as correlated.
In addition, an uncorrelated uncertainty on the polar angle measurement of the
scattered lepton of $2$~mrad for $\theta_{e} > 120^{\circ}$ and $3$~mrad elsewhere
is also included.


An uncertainty of $2$\% is assigned to the scale of the measured hadronic
energy for events in the CC event samples, of which $1$\% is considered to be a
correlated component.
In addition, a $10$\% correlated uncertainty is assigned to the amount of energy
in the LAr calorimeter attributed to noise for events in the CC event samples.


Other experimental systematic uncertainties are found to have a negligible
impact on the analysis.
The effect of the above systematic uncertainties on the SM expectation is determined
by varying the experimental quantities by $\pm 1$ standard deviation in the MC
samples and propagating these variations through the whole analysis.


The luminosity measurement has an average uncertainty of $3$\%.
The uncertainty on the polarisation measurement is $3.5$\% and is found to have a
negligible effect on the limit calculations performed in section~\ref{sec:limits}.


All data sets are compared to a SM prediction based on the
CTEQ6m~\cite{Pumplin:2002vw} parameterisation of the parton densities inside
the proton.
The uncertainties of this parameterisation are propagated through
the analysis using the full set of eigenvector PDFs, and the effect
is added in quadrature to the experimental uncertainties listed above.

\section{Results}
\label{sec:results}

\subsection{Mass distributions}
\label{sec:massspectra}

\begin{figure}[ht] 
  \begin{center}
   \includegraphics[width=0.96\textwidth]{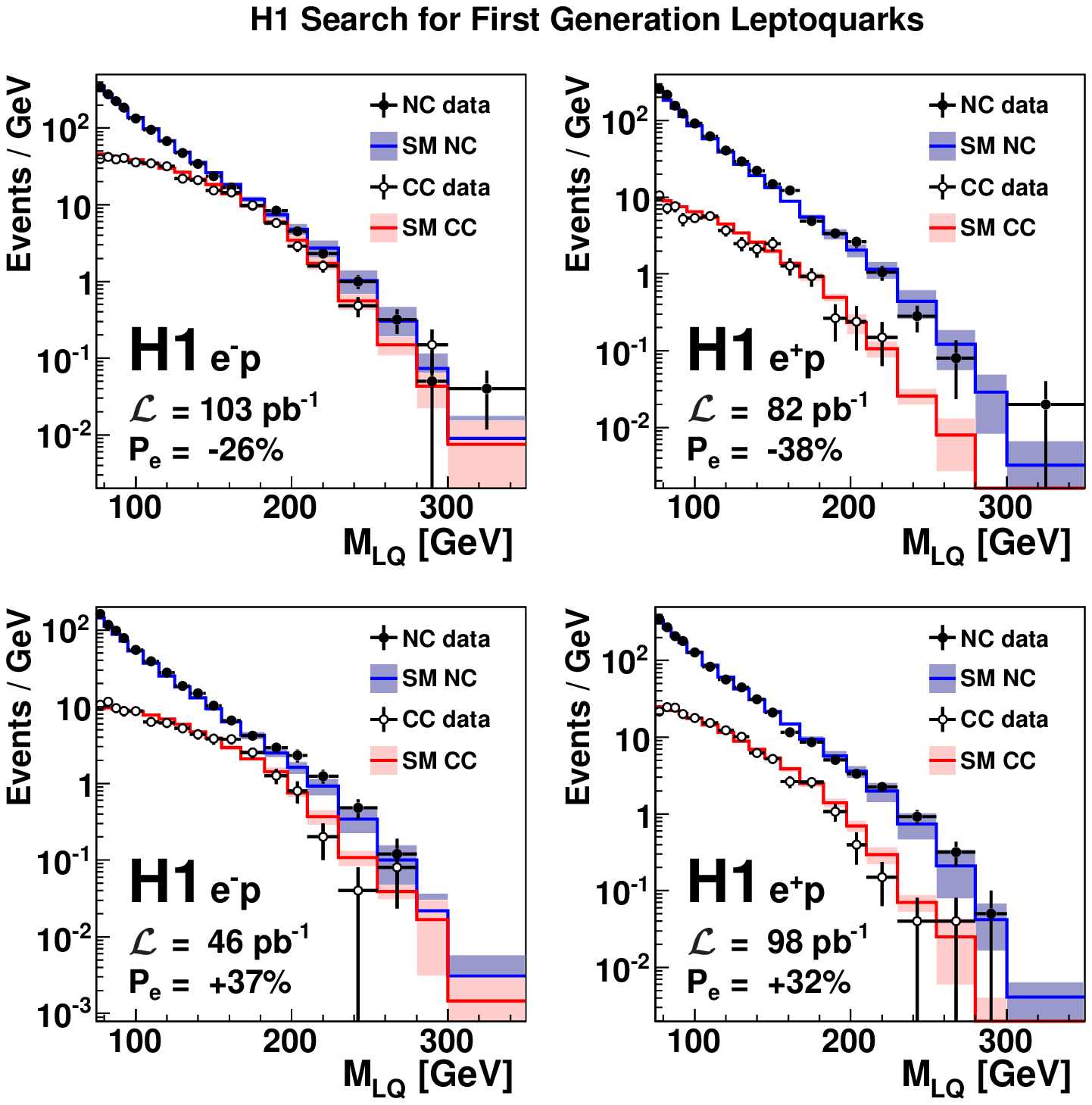}     
  \end{center} 
  \begin{picture} (0.,0.)
  \setlength{\unitlength}{1.0cm}
    \put (6.9,11.5){\bf\normalsize (a)}
    \put (14.6,11.5){\bf\normalsize (b)}
    \put (6.9, 4.3){\bf\normalsize (c)}
    \put (14.6,4.3){\bf\normalsize (d)}
 \end{picture}
   \caption{The reconstructed leptoquark mass in the search for first generation leptoquarks
      in the 2003-2007 H1 data, which was taken with a polarised lepton beam. The left-handed electron data (a)
      and left-handed positron data (b) are shown in the top row; the right-handed electron data (c) and
      right-handed positron data (d) are shown in the bottom row.
      The luminosity $\mathcal{L}$ and average longitudinal lepton polarisation $P_{e}$ of each data set is indicated.
      The NC (solid points) and CC (open points) data are compared to the SM predictions
      (histograms), where the shaded bands indicate the total SM uncertainties.}
 \label{fig:polmassplots}
\end{figure}

\begin{figure}[ht] 
  \begin{center}
   \includegraphics[width=0.96\textwidth]{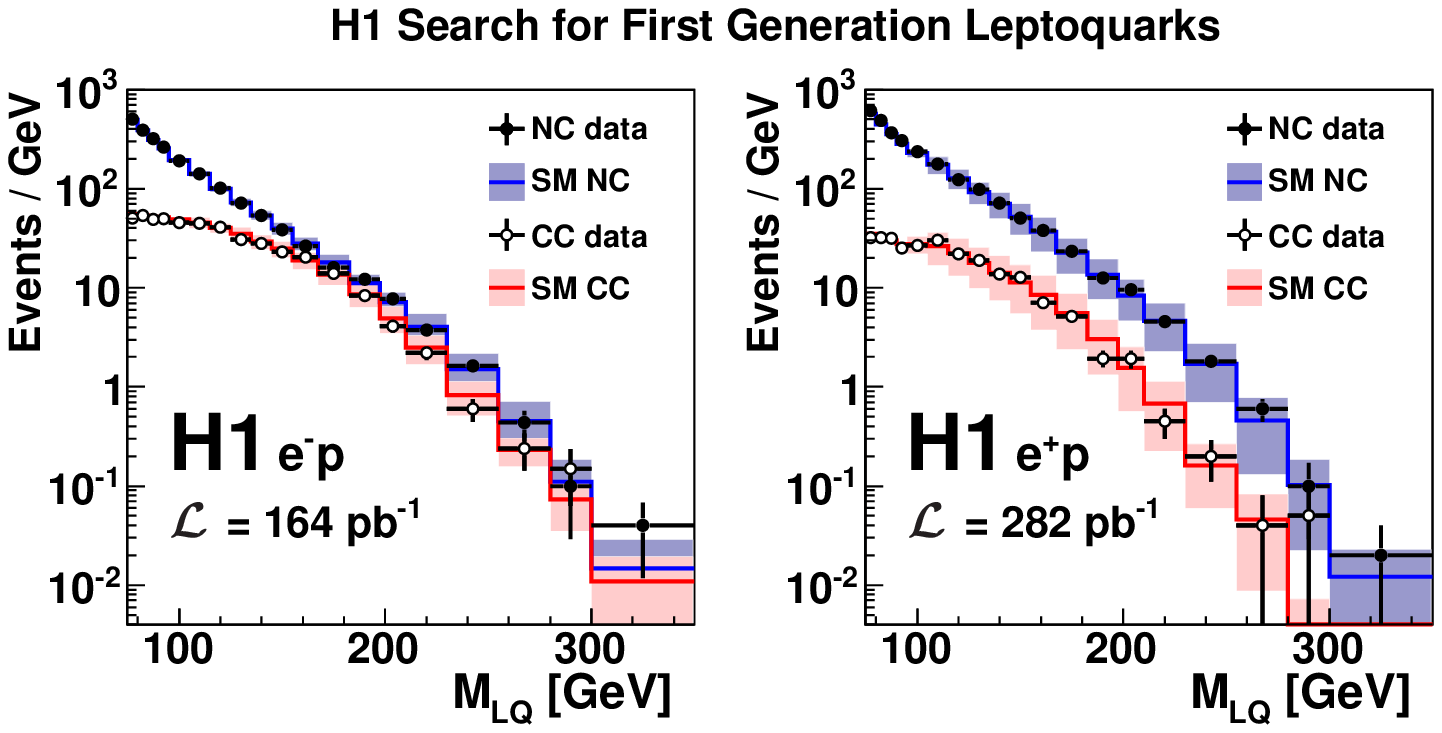}     
  \end{center}   
  \begin{picture} (0.,0.)
  \setlength{\unitlength}{1.0cm}
    \put (6.9,4.3){\bf\normalsize (a)}
    \put (14.6,4.3){\bf\normalsize (b)}
\end{picture}
    \caption{The reconstructed leptoquark mass in the search for first generation leptoquarks
      in the full H1 electron (a) and positron (b) data.
      The luminosity $\mathcal{L}$ of each data set is indicated.
      The NC (solid points) and CC (open points) data are compared to the SM predictions
      (histograms), where the shaded bands indicate the total SM uncertainties.}
\label{fig:fullmassplots}
\end{figure}

Mass spectra of the four H1 data sets taken with a longitudinally polarised lepton
beam as defined in table~\ref{tab:alldatasets} are shown in figure~\ref{fig:polmassplots},
where both the NC and CC event samples are presented.
The mass spectra of the complete electron and positron H1 data sets are presented
in figure~\ref{fig:fullmassplots}.
A good description of the H1 data by the SM is observed, where the
expectation is dominated by DIS processes in all event samples, with small
additional contributions from photoproduction.
Since no evidence for LQ production is observed in any of the NC or CC data samples,
the data are used to set constraints on LQs coupling to first generation fermions.

\subsection{Statistical Method}
\label{sec:method}

For the limit analysis, the data are studied in bins in the $M_{LQ}-y$ plane, where the
NC and CC data samples with different lepton beam charge and polarisation are 
kept as distinct data sets.
In total, $N_{\rm bin} = 1408$ bins are considered, divided equally between the NC and
CC event samples.
For a given bin $i$, the predicted LQ signal contribution is denoted $s_i$ and the predicted
number of events in the absence of a LQ signal is denoted $b_i$.
The number of events in the presence of a LQ signal is thus $s_i+b_i$ and is obtained
as a function of the LQ mass and coupling by a reweighting procedure~\cite{Aktas:2005pr}.
The limits are determined from a statistical analysis which uses the method of fractional event 
counting, optimised for the presence of systematic uncertainties~\cite{Bock:2004xz}. 
For a given leptoquark type, mass and coupling hypothesis, a test statistic
$X$ is constructed as a fractional event count of all events:
\begin{equation} 
  X \,\, = \,\, \sum_{i=1}^{N_{\rm bin}} w_i n_i\,,
\label{eqn:X}
\end{equation}
where the sum runs over all bins and $n_i$ is the number of events observed in bin $i$.
The weights $w_i$ are chosen such that in the presence of a LQ signal the test
statistic $X$ is larger than that expected from the SM.
In particular, bins with a large and positive $s_i$ have weights close to one,
whereas bins with $s_i$ close to zero have weights close to zero.
If $s_i$ is negative in a given bin due to interference effects,
the corresponding bin weight is also negative.
This has the desired effect that an event deficit in such a bin still leads to a $X$
larger than the SM expectation and thus is interpreted correctly as a signal
contribution.
The presence of systematic uncertainties may reduce the sensitivity of a given bin.
The weight is therefore defined in such a way as to ensure that only bins with both a
large signal contribution and small systematic uncertainties enter with sizeable weights
into the test statistic $X$.
This is achieved by defining the weights as solutions of the following set of
linear equations~\cite{Bock:2004xz}:
\begin{equation}
s_i  = k_1\left[(s_i+b_i)w_i+\sum_j V^{SB}_{ij}w_j\right]+k_2 \left[b_i w_i +\sum_j V^{B}_{ij}w_j\right]\,.
\label{eqn:si}
\end{equation}
In this analysis, the constants $k_1$ and $k_2$ are set to one, which is the appropriate choice
for testing signals with a well defined cross section prediction \cite{Bock:2004xz}.
The covariance matrices of all bins in the presence (absence) of the LQ signal,
$V^{SB}_{ij}$ ($V^{B}_{ij}$), are calculated as:
\begin{equation}
V^{SB}_{ij} = \sum_{k=1}^{N_{\rm sys}} \sigma^{SB}_{ki} \sigma^{SB}_{kj} \,\,\,\,\,\, \text {and} \,\,\,\,\,\,
V^{B}_{ij} = \sum_{k=1}^{N_{\rm sys}} \sigma^{B}_{ki} \sigma^{B}_{kj}\,,
\label{eqn:comat}
\end{equation}
where $\sigma^{SB}_{ki}$ ($\sigma^{B}_{ki}$) are the one sigma shifts induced from systematic
source $k$ to the number of events expected in bin $i$ in the presence (absence) of the
LQ signal.
The sums run over the $N_{\rm sys}$ sources of systematic uncertainty.
In the case of negligible systematic uncertainties, equation~\ref{eqn:comat} is
equivalent to the weight definition used in~\cite{Aktas:2005pr}. 


Limits are obtained by performing a frequentist analysis of the test statistic obtained from
the data, $X^{\rm data}$.
For each signal hypothesis, a large number of MC experiments, typically ${\cal O}(10^5)$,
are generated by varying the expected number of events $s_i+b_i$ within the uncertainties.
Systematic uncertainties are treated as Gaussian distributions and statistical fluctuations
are simulated using Poisson statistics.
For each MC experiment $e$, a test statistic $X^e$ is calculated.
A probability $p^{\rm data}$ is calculated as the fraction of MC experiments which
have $X^e<X^{\rm  data}$.
The LQ hypothesis is excluded at a given confidence level (CL) if $p^{\rm data}<1-$CL.
%
%
In addition to this condition, a power constraint is applied \cite{Cowan:2011an}.
The power constraint avoids the exclusion of LQ signals beyond the sensitivity of the
experiment, which may otherwise occur due to statistical fluctuations in the data in the
opposite direction to that expected from the LQ hypothesis.
A probability $p^{1\sigma}$ is determined as the fraction of MC experiments
with $X^e<X^{1\sigma}$.
Here, $X^{1\sigma}$ is the value of the test statistic which corresponds to
a $1\sigma$ downwards fluctuation of the SM.
It is determined from a second set of MC experiments, where each
experiment $\tilde{e}$ is simulated in the absence of a LQ signal,
that is by simulating systematic and statistical fluctuations of $b_i$.
The value $X^{1\sigma}$ is determined such that the fraction of MC experiments with
$X^{\tilde{e}}<X^{1\sigma}$ is equal to the single-sided $1\sigma$ quantile, $15.9$\%.
LQ models are excluded at $95\%$ CL with the power constraint applied,
if both $p^{\rm data}$ and $p^{1\sigma}$ are below $0.05$.

\subsection{Limits}
\label{sec:limits}

\begin{figure}[p] 
\begin{center}
      \includegraphics[width=.49\textwidth]{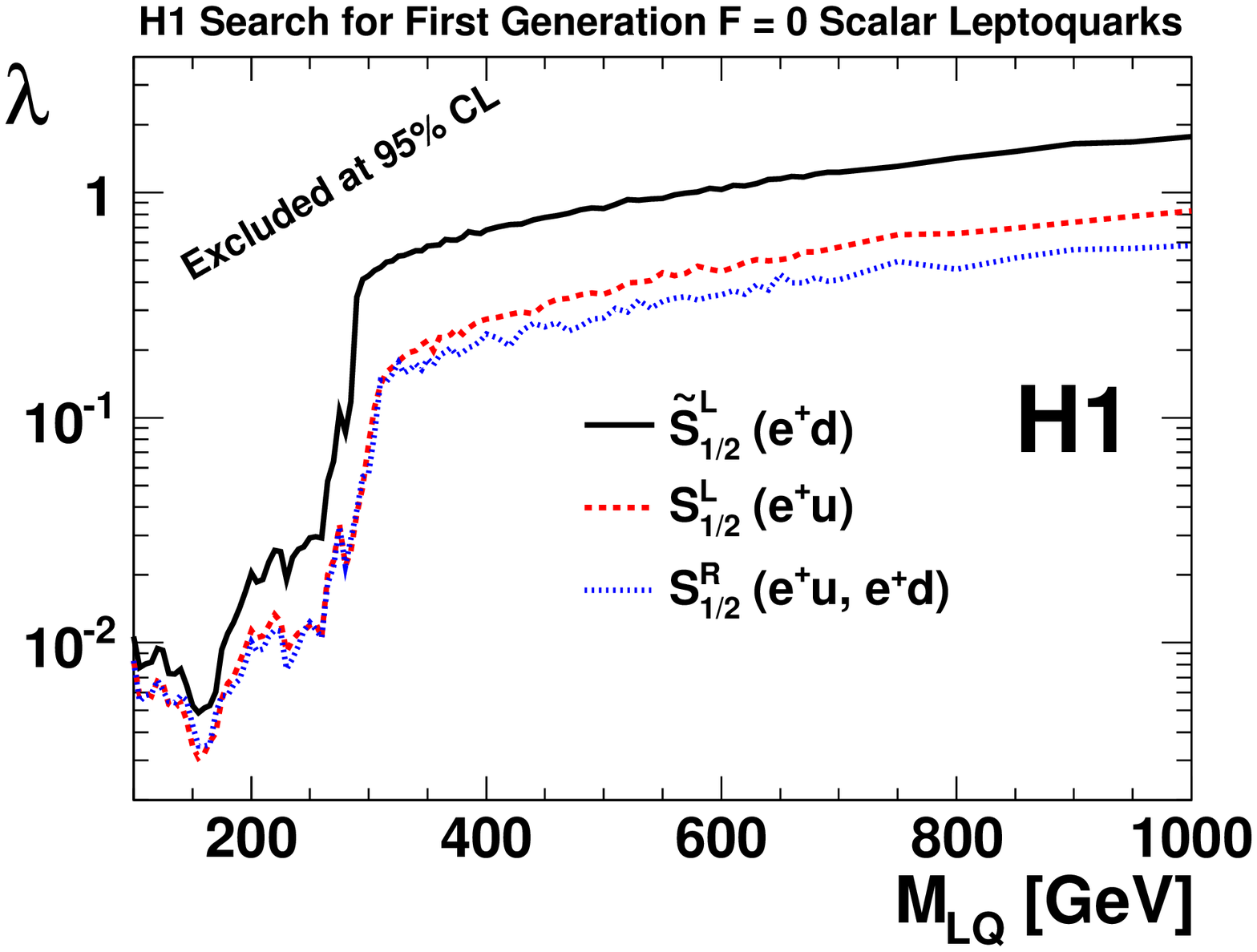}
      \includegraphics[width=.49\textwidth]{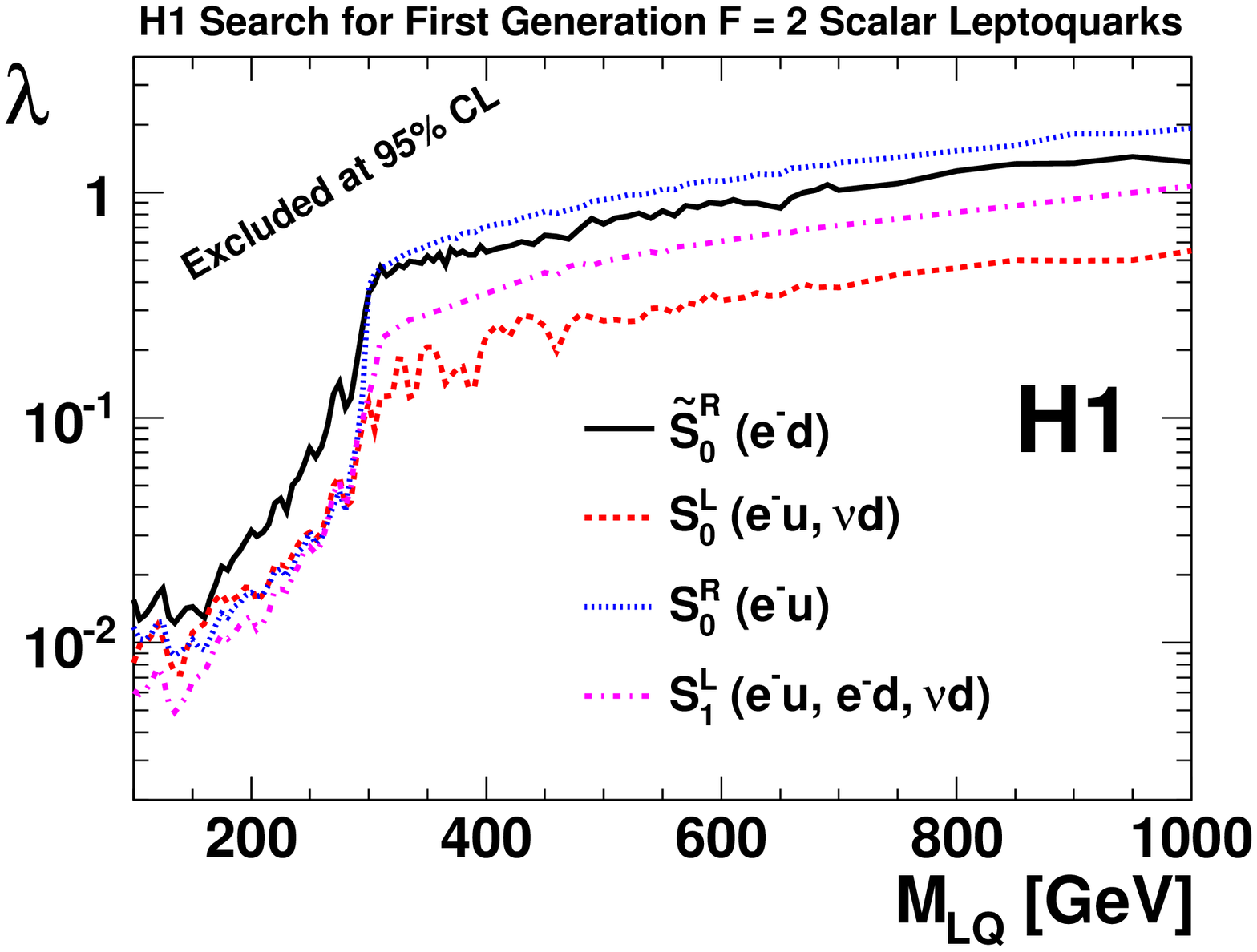}
      \includegraphics[width=.49\textwidth]{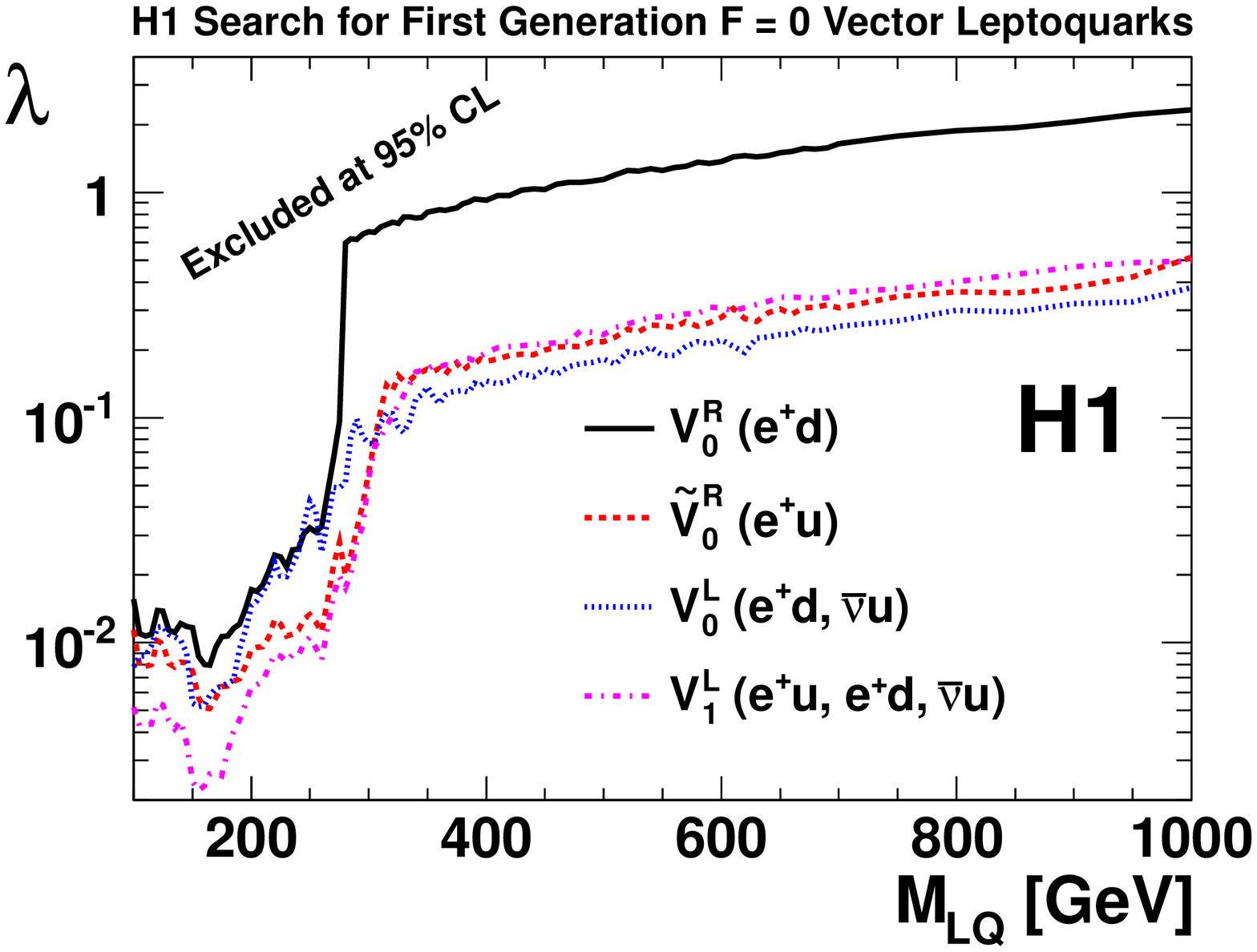}
      \includegraphics[width=.49\textwidth]{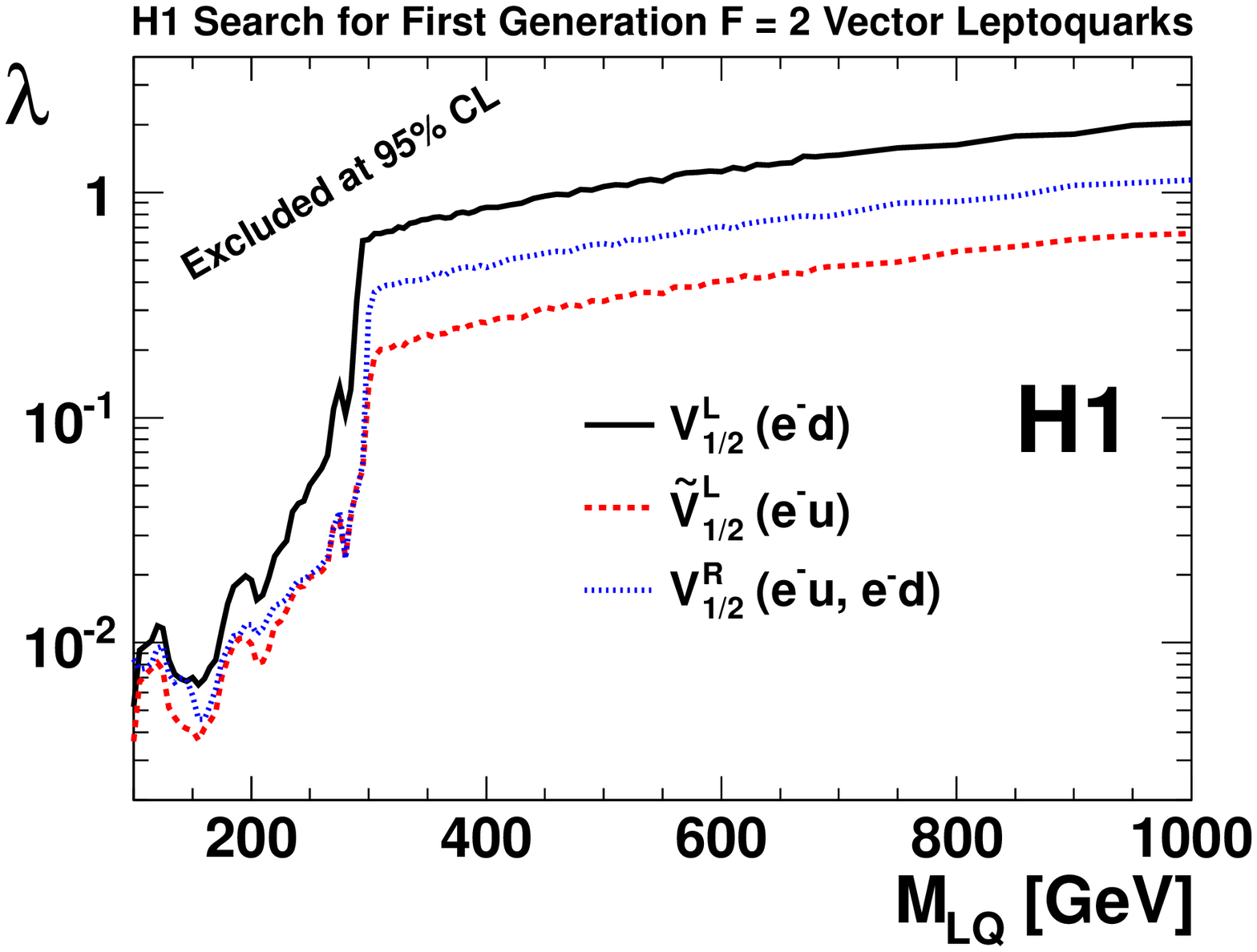}
\end{center}
  \begin{picture} (0.,0.)
  \setlength{\unitlength}{1.0cm}
    \put (2.6,8.2){\bf\normalsize (a)}
    \put (10.5,8.2){\bf\normalsize (b)}
    \put (2.6,2.3){\bf\normalsize (c)}
    \put (10.5,2.3){\bf\normalsize (d)}
 \end{picture}
  \caption{Exclusion limits for the 14 leptoquarks (LQs) described by the Buchm\"uller, R\"uckl and
    Wyler (BRW) model. The limits are expressed on the coupling $\lambda$ as a function of leptoquark
    mass for the scalar LQs with $F = 0$ (a) and $F = 2$ (b) and the vector LQs with $F = 0$ (c) and $F = 2$ (d).
    Domains above the curves are excluded at $95\%\,{\rm CL}$. The parentheses after the LQ name indicate the
    fermion pairs coupling to the LQ, where pairs involving anti-quarks are not shown.}
\label{fig:limits}
 \end{figure}

Exclusion limits are first derived within the phenomenological BRW model~\cite{BRW}
described in section~\ref{sec:lqs}.
Upper limits on the coupling $\lambda$ obtained at $95$\%~CL are
shown as a function of the leptoquark mass in figure~\ref{fig:limits}, displayed as
groups of scalar and vector LQs for both $F = 2$ and $F = 0$.
The presented limits extend beyond those from previous leptoquark and contact interaction
analyses based on smaller HERA data sets by the H1~\cite{Aktas:2005pr,Adloff:2003jm} and
ZEUS~\cite{Chekanov:2003af,Chekanov:2003pw} collaborations.
For a coupling of electromagnetic strength $\lambda = \sqrt{4 \pi \alpha_{\rm em}} = 0.3$,
LQs produced in $ep$ collisions decaying to an electron-quark or a neutrino-quark pair are
excluded at $95\%$~CL up to leptoquark masses between $277$~GeV~($V_{0}^{R}$)
and $800$~GeV~($V_{0}^{L}$), depending on the leptoquark type.

\begin{figure}[ht] 
  \begin{center}
    \includegraphics[width=0.74\textwidth]{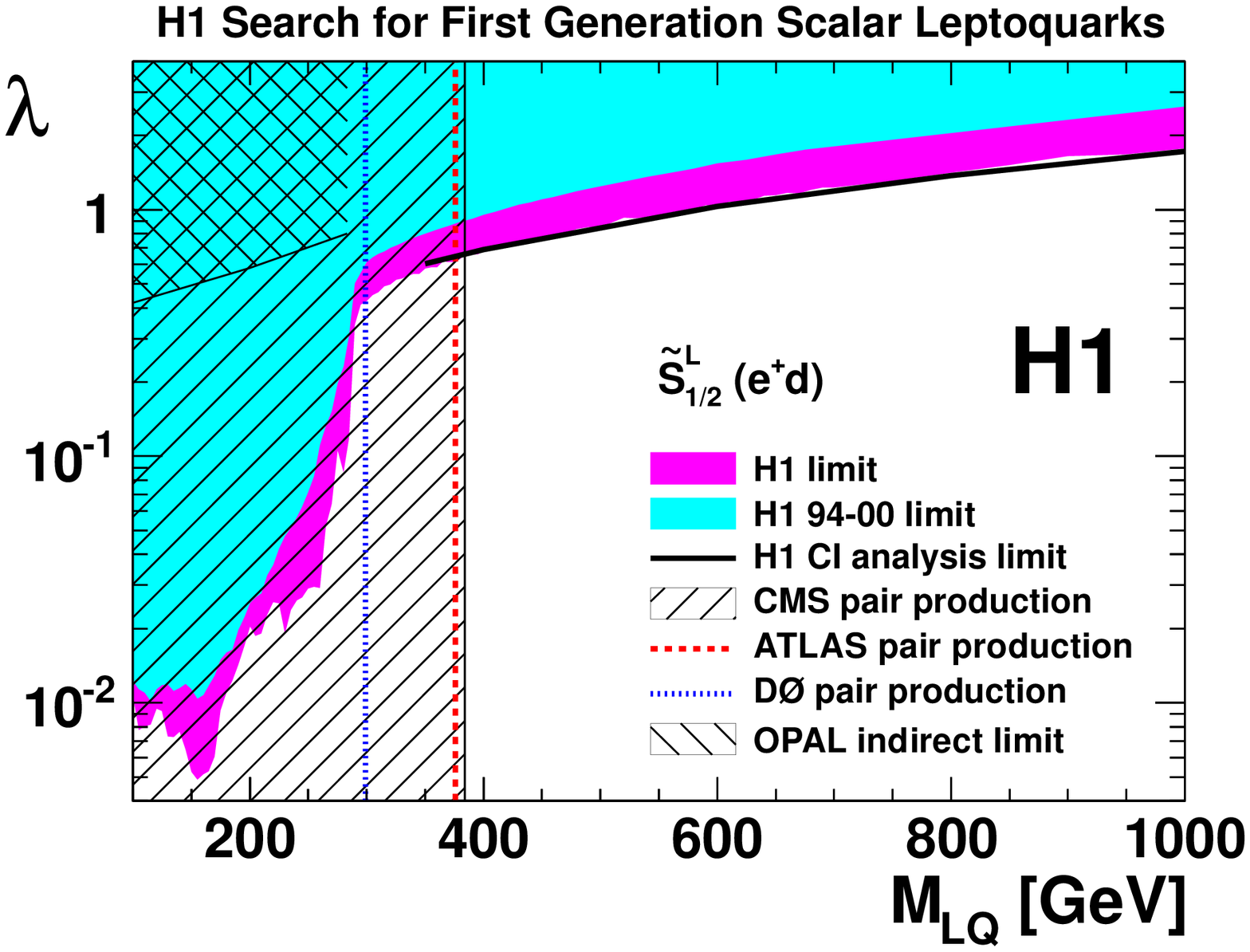}
    \includegraphics[width=0.74\textwidth]{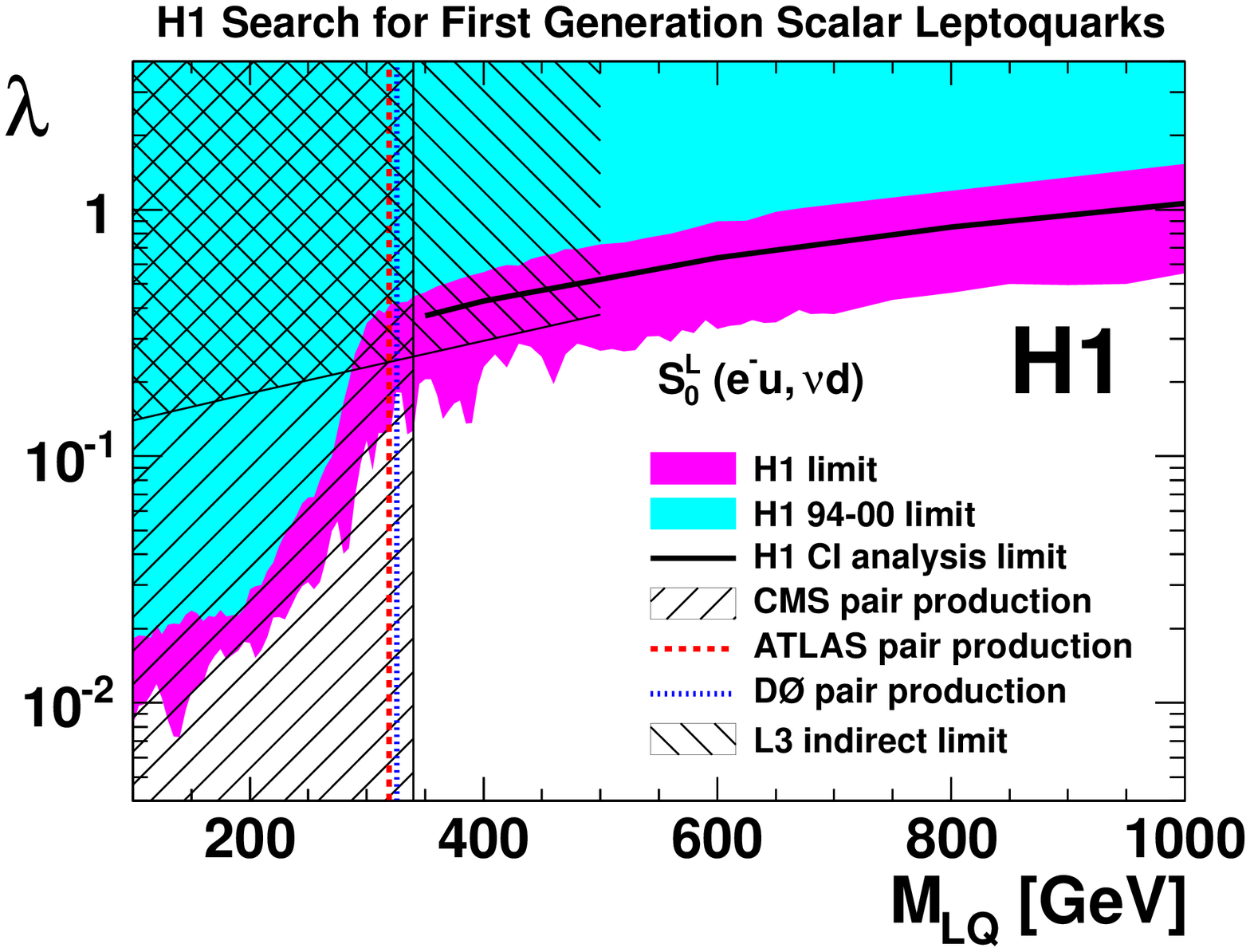}     
  \end{center} 
  \begin{picture} (0.,0.)
  \setlength{\unitlength}{1.0cm}
    \put (7.4,12.9){\bf\normalsize (a)}
    \put (7.4,4){\bf\normalsize (b)}
  \end{picture}
   \caption{Exclusion limits on the coupling $\lambda$ as a function of the leptoquark
     mass for the $\tilde{S}_{1/2}^{L}$ (a) and $S_{0}^{L}$ (b) leptoquarks in the framework of the
     BRW model. The parentheses after the LQ name indicate the
     fermion pairs coupling to the LQ, where pairs involving anti-quarks are not shown.
     Domains above the curves are excluded at $95$\%~CL.
     Limits from the previous H1 publication (94-00) are also indicated.
     For comparison, limits from LEP (OPAL and L3), the Tevatron (D{\O}) and the
     LHC (CMS and ATLAS, $\sqrt{s} =7$ TeV data) are shown for comparison, as well as constraints
     on LQs with masses above $350$~GeV from the H1 contact interaction (CI) analysis.}
 \label{fig:compare}
\end{figure}

Within the framework of the BRW model, the $\tilde{S}_{1/2}^{L}$ LQ decays exclusively to an
electron-quark pair, resulting in a branching fraction for decays
into charged leptons of $\beta_{e} = 1.0$, whereas the $S_{0}^{L}$ LQ also decays
to neutrino-quark, resulting in $\beta_{e} = 0.5$.
The H1 limits on $\tilde{S}_{1/2}^{L}$ and $S_{0}^{L}$ presented in this paper are compared
to those from other experiments in figure~\ref{fig:compare}.
Limits from the previous H1 publication~\cite{Aktas:2005pr} are also shown.
Indirect limits from searches for new physics in $e^+e^-$ collisions at LEP by the
OPAL~\cite{Abbiendi:1998ea} and L3~\cite{Acciarri:2000uh} experiments
are indicated, as well as the limits from D{\O}~\cite{Abazov:2009gf,Abazov:2011qj} at the Tevatron
and from the  CMS~\cite{Khachatryan:2010mp,Chatrchyan:2011ar}  and
ATLAS~\cite{Aad:2011uv} experiments at the LHC. 
The limits from hadron colliders are based on searches for LQ pair-production and are
independent of the coupling $\lambda$, where the strongest current limit for
$\beta_{e} = 1.0$ ($\beta_{e} = 0.5$) scalar LQs is $384$~GeV ($340$~GeV)
as reported by the CMS collaboration.
For these leptoquark masses, this analysis rules out the $\tilde{S}_{1/2}^{L}$ and
$S_{0}^{L}$ LQs for coupling strengths larger than $0.64$ and $0.14$ respectively.
The H1 limits at high leptoquark mass values are also compared with those
obtained in a contact interaction analysis~\cite{newh1CI}, which is based on
single differential NC cross sections ${\rm d}\sigma/{\rm d}Q^2$ measured
using the same data.
The additional impact of the CC data can be seen in the case of the $S_{0}^{L}$ LQ,
where a stronger limit is achieved in this analysis, whereas for the $\tilde{S}_{1/2}^{L}$
LQ the two analyses result in a similar limit.


Signatures similar to those expected from LQ decays also appear in
supersymmetric models with $R$-parity violation~\cite{Butterworth:1992tc}.
In such models, the production and direct decay of the $\tilde{u}^{j}_{L}$
($\tilde{d}^{k}_{R}$) squark via a $\lambda'_{1j1}$ ($\lambda'_{11k}$) coupling is equivalent
to the interaction of the $\tilde{S}_{1/2}^{L}$ ($S_{0}^{L}$) LQ with a lepton-quark pair,
and as such the results described in the previous paragraph are also valid for these squark types,
assuming the direct decay dominates.
More general limits on squark production with $R$-parity violating decays are
presented in a dedicated H1 publication~\cite{newh1SUSY}.

\begin{figure}[ht] 
  \begin{center}
    \includegraphics[width=0.74\textwidth]{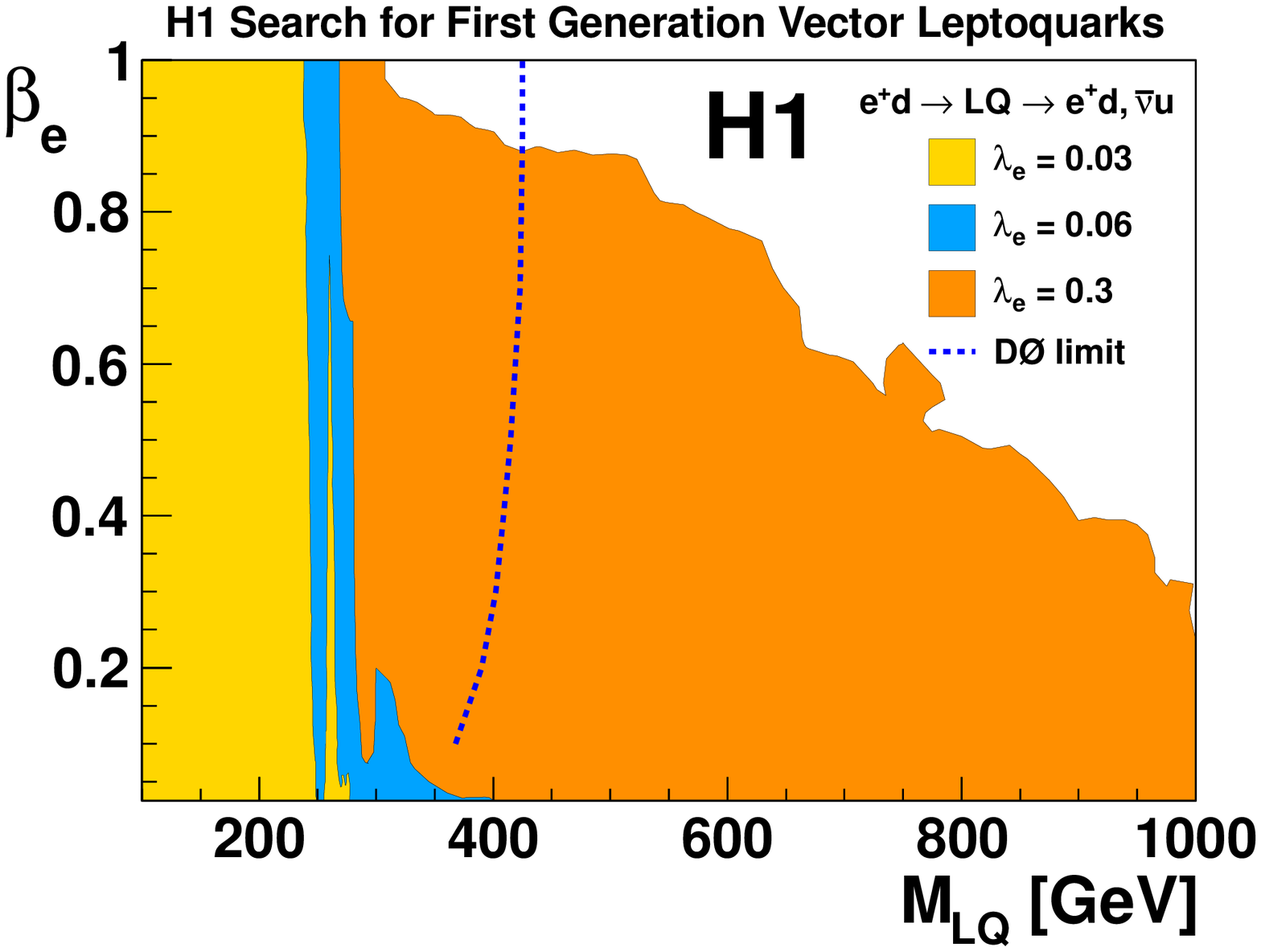}     
    \includegraphics[width=0.74\textwidth]{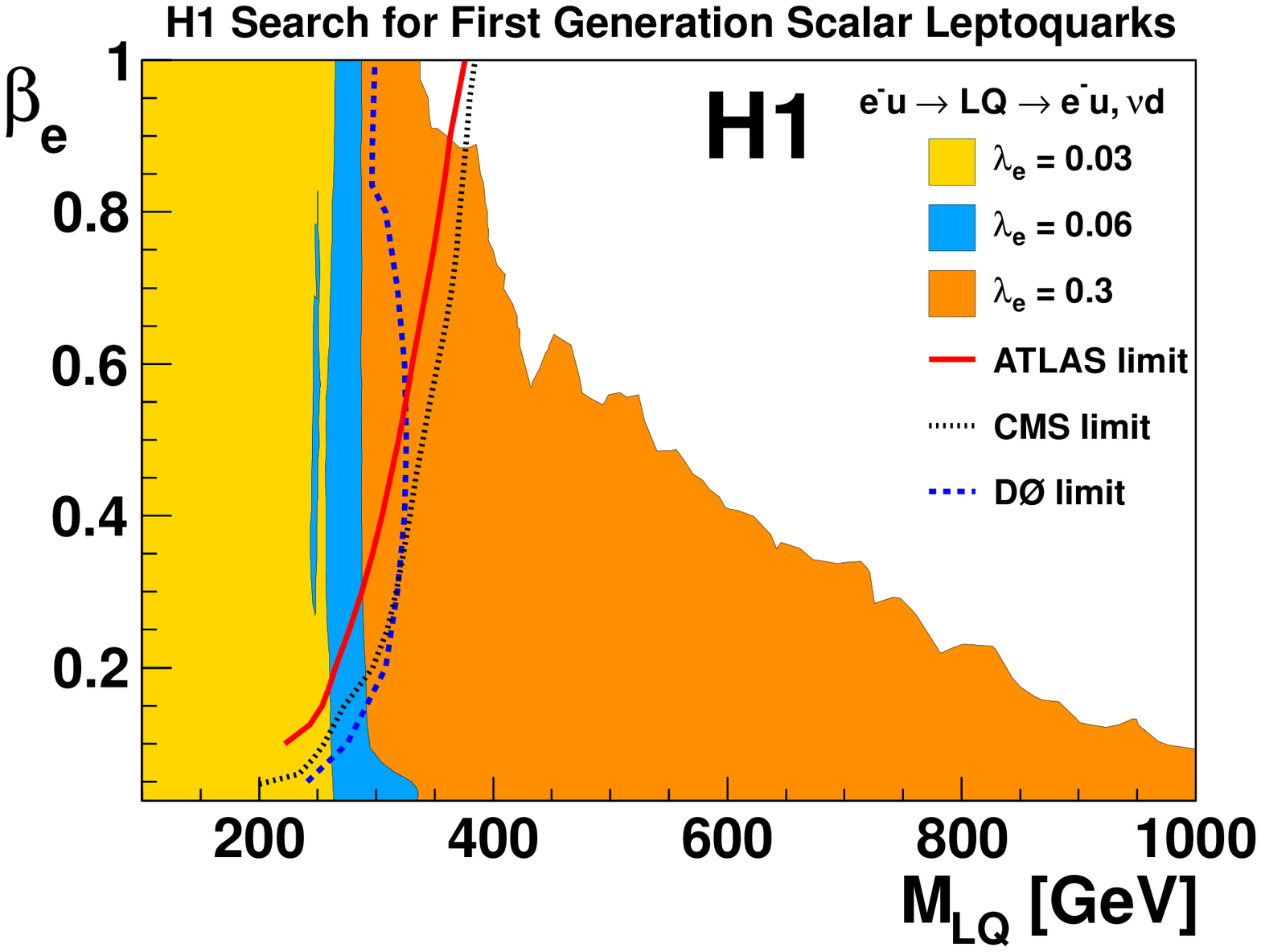}     
  \end{center} 
  \begin{picture} (0.,0.)
  \setlength{\unitlength}{1.0cm}
    \put (12.0,12.0){\bf\normalsize (a)}
    \put (12.0,3.8){\bf\normalsize (b)}
  \end{picture}
  \caption{Regions of $\beta_e$--$M_{\rm LQ}$ ruled out by the combination
    of the NC and CC analyses for (a) a vector LQ coupling to $e^+d$ (with the
    quantum numbers of the $V_{0}^{L}$) and (b) for a scalar LQ coupling to
    $e^-u$ (with the quantum numbers of the $S_{0}^{L}$),  where only LQ decays
    into $eq$ and $\nu q$ are considered.
    Excluded regions at $95$\%~CL are shown as the coloured areas for
    three different values of the electron-quark-LQ coupling $\lambda_{e}$.
    Limits from the Tevatron (D{\O}) and the LHC (CMS and ATLAS, $\sqrt{s} =7$ TeV data),
    which do not depend on $\lambda_{e}$, are also indicated.}
  \label{fig:betascan}
\end{figure}

Beyond the BRW ansatz, $\beta_{e}$ may be considered as a free parameter and the
couplings and therefore also the branching ratios to electron-quark and neutrino-quark
are not necessarily equal.
By investigating such a model, mass dependent constraints on the LQ branching
ratio $\beta_{e}$ can be set for a given value of the electron-quark-LQ
coupling $\lambda_{e}$.
Excluded regions in the $\beta_{e}$--$M_{\rm LQ}$ plane for three different coupling
strengths are shown for a vector LQ with quantum numbers identical to $V_{0}^{L}$ in
figure~\ref{fig:betascan}(a) and for a scalar LQ with quantum numbers identical
to $S_{0}^{L}$ in figure~\ref{fig:betascan}(b).
Similar exclusion limits from the Tevatron (D{\O}~\cite{Abazov:2009gf,Abazov:2011qj}) and the LHC
(CMS~\cite{Chatrchyan:2011ar} and ATLAS~\cite{Aad:2011uv}), which do not depend
on $\lambda_{e}$, are also shown in figure~\ref{fig:betascan}.
For an electron-quark-LQ coupling of electromagnetic strength $\lambda_{e}=0.3$ the
H1 limits extend to high leptoquark masses beyond the kinematic limit of resonant LQ
production, and for most values of $\beta_{e}$ extend considerably beyond the region
currently excluded by hadron colliders.

\section{Summary}
\label{sec:summary}

A search for first generation scalar and vector leptoquarks is performed using the complete
H1 $e^{\pm}p$ data taken at a centre-of-mass energy of up to $319\,{\rm GeV}$ and 
corresponding to an integrated luminosity of $446$~pb$^{-1}$.
The H1 data are well described by the SM prediction and no leptoquark signal is observed.
Limits are derived on $14$ leptoquark types and assuming a coupling strength of
$\lambda=0.3$ leptoquarks are ruled out up to masses of $800$~GeV, which is
beyond the current limits from hadron colliders.

\section*{Acknowledgements}

We are grateful to the HERA machine group whose outstanding efforts have made this experiment
possible. We thank the engineers and technicians for their work in constructing and maintaining the
H1 detector, our funding agencies for financial support, the DESY technical staff for continual
assistance and the DESY directorate for support and for the hospitality which they extend to the
non-DESY members of the collaboration.



\end{document}